\documentclass[proceedings,nohyper]{JHEP} 

\usepackage{epsfig}			

\newbox\mybox
\newcommand\fverb{\setbox\mybox=\hbox\bgroup\verb}
\newcommand\fverbdo{\egroup\medskip\noindent\fbox{\unhbox\mybox}\ }
\newcommand\fverbit{\egroup\item[\fbox{\unhbox\mybox}]}

\font\beeg=cmr17 scaled 1600		
\newcommand\init[1]{\setbox\mybox=\hbox{{\beeg #1}~}%
		   \noindent\global\hangindent=\wd\mybox\global\hangafter-2%
		   \sc\smash{\llap {\lower 13.2pt \box\mybox}}}

\newcommand{\pha}{\phi_1}
\newcommand{\phit}{\tilde{\phi}}

\newcommand{\Det}{{\rm Det}}
\newcommand{\phb}{\phi_2}

\newcommand{\lx}{\lambda}
\newcommand{\Lx}{\Lambda}

\newcommand{\kx}{\kappa}
\newcommand{\be}{\begin{equation}}
\newcommand{\ee}{\end{equation}}
\newcommand{\een}{\end{subequations}}
\newcommand{\ben}{\begin{subequations}}
\newcommand{\beq}{\begin{eqnarray}}
\newcommand{\eeq}{\end{eqnarray}}
\def \lta {\mathrel{\vcenter
     {\hbox{$<$}\nointerlineskip\hbox{$\sim$}}}}
\def \gta {\mathrel{\vcenter
     {\hbox{$>$}\nointerlineskip\hbox{$\sim$}}}}

\title{A consistent approach to bubble-nucleation theory}

\author{N. Tetradis\thanks{Report on 
	work done in collaboration with A. Strumia and
	C. Wetterich.}\\
	Scuola Normale Superiore, Piazza dei Cavalieri 7, Pisa 56126, Italy\\ 
	E-mail: \email{tetradis@cibs.sns.it}}

\conference{Trieste Meeting of the TMR Network on Physics beyond the SM}

\abstract{We summarize recent work on the consistent 
calculation of bubble-nucleation rates. Our approach is based on the
notion of a real coarse-grained potential. The bubble-nucleation
rate is calculated through
an expansion around the semiclassical saddle point associated with
tunnelling. We resolve outstanding problems related to the convexity
of the potential, the double-counting of the effect of fluctuations and
the inherent ultraviolet divergences.  
We determine the region of validity of the expansion around the saddle point.
We find that this expansion fails near the spinodal line, and for weak
or radiatively induced first-order phase transitions.
We apply our method to the bound on the Higgs-boson mass from vacuum
metastability and the electroweak phase transition.
}

\begin{document} 

\maketitle 


{\init Developing} a consistent theory of bubble nucleation is 
an important problem for particle and statistical physics.
Possible applications include the description of most cosmological phase
transitions, the questions of tunnelling in quantum mechanics and
field theory, the problem of vacuum stability 
and its phenomenological consequences, the (currently unsatisfactory)
explanation of the experimentally
measured bubble-nucleation rates for statistical systems, etc.  
The estimates of bubble-nucleation ra\-tes for
first-order phase transitions are usually carried out 
within Langer's theory of homogeneous nucleation~\cite{langer}, 
applied to relativistic field theory in refs.~\cite{coleman}.
The nucleation rate is exponentially suppressed by the 
action (free energy rescaled by the temperature) of the
critical bubble, 
a saddle point of the free energy of the system.
Significant contributions to the nucleation rate 
may arise from higher orders in a systematic expansion around
this saddle point. The first correction has the form of a pre-exponential
factor that involves fluctuation determinants around the critical bubble 
and the false vacuum. 
The evaluation of this factor is a difficult
problem at the conceptual and technical level, as crucial issues associated
with the convexity of the potential, the divergences of the fluctuation 
determinants 
and the double-counting of the effect of fluctuations 
must be resolved.

\section{The method}
In a series of recent works~\cite{first}--\cite{fifth}, following 
the proposal of refs.~\cite{bubble1}, we 
developed a consistent approach, based on the effective average action
$\Gamma_k$~\cite{averact} that
can be interpreted as 
a coarse-grained free energy.
Fluctuations with 
characteristic momenta larger than a coarse-graining scale
($q^2 \gta k^2$) are integrated out
and their effect is incorporated in 
$\Gamma_k$. 
In the limit $k \to 0$,
$\Gamma_k$ becomes the effective action.
The $k$ dependence 
of $\Gamma_k$ is described by an exact flow equation~\cite{exact}.
This can be translated into evolution equations
for functions appearing in a derivative expansion of
the action~\cite{indices}. An approximation that is sufficient 
in many cases  
takes into account the effective average potential
$U_k$ and a standard kinetic term and neglects higher derivative
terms in the action.  
The bare theory is defined
at some high scale $\Lx$ that can be identified with the ultraviolet 
cutoff. 
It is, however, more convenient to 
choose a starting scale $k_0$ below the temperature $T$,
where the effective average action of
a $(3+1)$-dimensional theory at non-zero temperature
can be described in terms of an effective  
three-dimensional action at zero temperature~\cite{trans}.

In ref.~\cite{first} we computed the form of $U_k$ at scales $k\leq k_0$ for 
a theory of one scalar field by
integrating its evolution equation~\cite{exact,indices}
\be
\frac{\partial}{\partial k^2} U_k(\phi) =
-\frac{1}{8 \pi} {\rm Tr} \sqrt{k^2 + U''_k(\phi)}.
\label{evpot} \ee
We considered 
an initial potential $U_{k_0}$ with two minima separated by a barrier.
$U_k$ is real and non-convex for non-zero $k$, and 
approaches convexity only in the limit $k\to 0$.
The nucleation rate must be computed for $k$ larger than the scale $k_f$
for which the negative curvature at the top
of the barrier becomes approximately equal to 
$-k_f^2$~\cite{convex1}. 
For $k<k_f$ the form of the potential
is affected by field configurations that interpolate between
the two minima (similar to critical bubbles). 
For $k \gta k_f$ the typical length scale of a thick-wall critical
bubble is $\gta  1/k$. 

We performed the calculation of the nucleation rate for a range of scales
above and near $k_f$.
The nucleation rate is given by 
$$I=A_{k} \exp({-S_k}),$$
where
$$A_{k}= \frac{E_0}{2\pi}\left(\frac{S_k}{2\pi}\right)^{3/2}
\bigg|
\frac{\Det'\left[-\partial^2+U''_{k}(\phi_b(r)) \right]}
{\Det \left[ -\partial^2+k^2 +U''_{k}(\phi_b(r))\right]}\times$$
\beq\times
\frac{\Det\left[-\partial^2+k^2+U''_{k}(0) \right]}
{\Det\left[-\partial^2+U''_{k}(0)\right]}
\bigg|^{-1/2}.
\label{rrate} \eeq
Here $\phi_b(r)$ is the profile of the spherically symmetric
saddle point, $S_k$ its action determined by using the potential
$U_k(\phi)$,
and $\phi = 0$ corresponds to the false vacuum. 
The prime in the fluctuation determinant around
the saddle point denotes that the 3 zero eigenvalues 
of the operator $-\partial^2+U''_{k}(\phi_b(r))$, corresponding to
displacements of the critical bubble, 
have been removed. $E_0$ is the square root of
the absolute value of the unique negative eigenvalue.
The above form of $A_k$ guarantees that only 
modes with characteristic momenta $q^2 \lta k^2$ contribute to the
nucleation rate. (The effect of modes with $q^2 \gta k^2$ is already 
incorporated in the form of $U_k(\phi)$.) We emphasize that the
use of an ultraviolet cutoff in eq.~(\ref{rrate}) that matches the
infrared cutoff procedure in the derivation of eq.~(\ref{evpot})
is crucial for the consistency of our approach. In both cases, mass-like
cutoffs have been used.

\section{Typical calculations}
The details of the numerical evaluation of the complicated determinants
in eq.~(\ref{rrate}) are given in ref. \cite{first}.
Fig.~\ref{fig1} exhibits the results of the calculation
for an initial potential 
\be
U_{k_0} (\phi) =
\frac{1}{2}m^2_{k_0} \phi^2
+\frac{1}{6} \gamma_{k_0} \phi^3
+\frac{1}{8} \lx_{k_0} \phi^4
\label{eq:two20} \ee
that has a form typical for
first-order phase transitions in four-dimensional 
field theories at high temperature.
Through a shift $\phi\to\phi+c$ the cubic term can be eliminated 
in favour of a term linear in $\phi$.
Therefore, eq. (\ref{eq:two20}) also
describes statistical systems of the Ising universality class
in the presence of an external magnetic field. 
The first row corresponds to parameters
$m^2_{k_0}=-0.0433~k_0^2$, 
$\gamma_{k_0}=-0.0634~k_0^{3/2}$,
$\lx_{k_0}=0.1~k_0$. The first plot shows
the evolution of the potential $U_k(\phi)$ as the scale
$k$ is lowered.
(We always shift the metastable vacuum to $\phi=0$.)~
The solid line corresponds to $k/k_0=0.513$ while the
line with longest dashes (that has the smallest barrier height)
corresponds to $k_f/k_0=0.223$. At the scale $k_f$ the negative
curvature at the top of the barrier is slightly larger than
$-k_f^2$ and we stop the evolution.
The potential and the field have been
normalized with respect to $k_f$.
As $k$ is lowered from $k_0$ to $k_f$, the absolute minimum of the potential
settles at a non-zero value of $\phi$, with a significant barrier
separating it from the metastable minimum at $\phi=0$.
The second plot displays the profile of the saddle point $\phi_b(r)$
in units of $k_f$
for the same sequence of scales.  For $k\simeq k_f$ the characteristic
length scale of the bubble profile and $1/k$ are comparable. 
This is an indication that we should not proceed to coarse-graining
scales below $k_f$.

\FIGURE{\epsfig{file=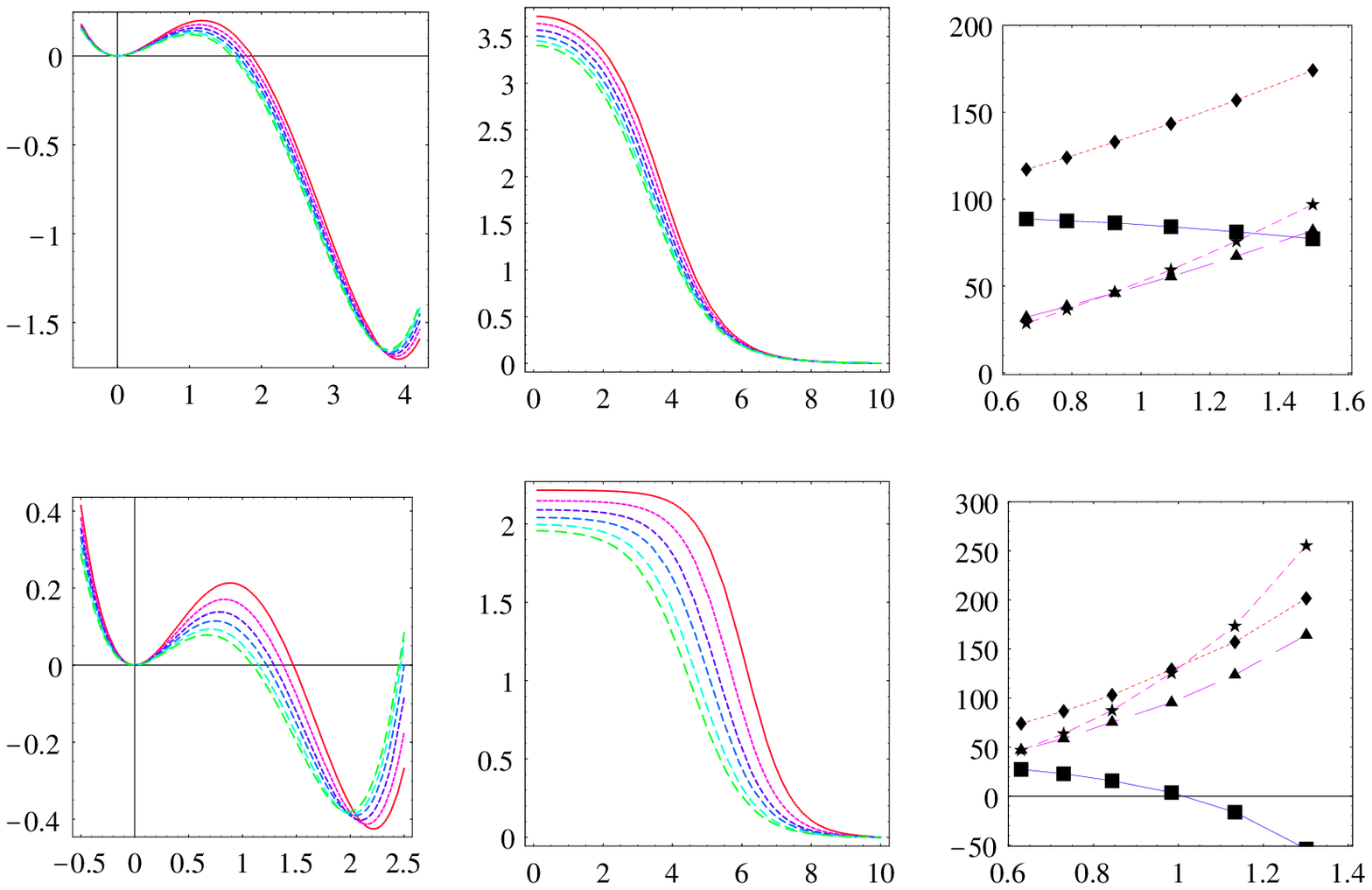,width=15.cm}%
        \caption{
Dependence of potential $U_k(\phi)$, critical bubble $\phi_b(r)$ and
nucleation rate $I/k^4_f$ on the coarse graining scale $k$.
The initial potential is given by eq.~(\ref{eq:two20}) 
with $\lambda_{k_0}=0.1\cdot k_0$,
$m^2_{k_0}=-0.0433\cdot k_0^2$, $\gamma_{k_0}=-0.0634~k_0^{3/2}$
(first row)
and $\lambda_{k_0}=0.1\cdot k_0$,
$m^2_{k_0}=-0.013\cdot k_0^2$, $\gamma_{k_0}=-1.61 \cdot 10^{-3}~k_0^{3/2}$
(second row).}
	\label{fig1}}

Our results for the nucleation rate are presented in the third plot.
The horizontal axis corresponds to $k/\sqrt{U''_k(\phi_t})$,
i.e. the ratio of the scale $k$
to the mass of the field at the
absolute minimum located at $\phi_t$.
Typically, when $k$ crosses below this mass, 
the massive fluctuations of the field
start decoupling. The evolution of the convex parts of
the  potential slows down and eventually stops.
The dark diamonds give the values of the action $S_k$ (free energy 
rescaled by the temperature)
of the saddle point at the scale $k$. We observe a strong
$k$ dependence of this quantity.
The stars indicate the values of
$\ln ( A_k/k^4_f )$.
Again a substantial decrease with decreasing $k$ is observed. This is expected,
because the fluctuation determinants in $A_k$ are calculated with an 
effective ultraviolet cutoff of order $k$.
The dark squares give our results for
$-\ln(I/k^4_f )
= S_k-\ln ( A_k/k^4_f )$. 
The
$k$ dependence of this quantity almost disappears for $k/\sqrt{U''_k(\phi_t})
\lta 1$.
The small residual dependence on $k$ can be used in order to estimate the
effect of the next order in the expansion around the saddle point.
This contribution is expected to be
smaller than $\ln ( A_k/k^4_f )$.
It is apparent from fig.~\ref{fig1}
that the leading contribution to the 
pre-exponential factor increases the total nucleation rate. 
This behaviour, associated with 
the fluctuations of the field whose expectation 
value serves as the order parameter, is observed
in multi-field models as well.

Our results confirm the expectation that the
nucleation rate should be independent of the scale $k$ that
we introduced as a calculational tool. They also demonstrate that
all the configurations of the second plot give equivalent
descriptions of the system, at least for the lower values of $k$.
This indicates that the critical bubble should not be associated only
with the saddle point of the semiclassical approximation, whose
action is scale dependent. Only the combination of
the saddle point and its possible deformations
in the thermal bath has physical meaning.

For larger values of 
$\lambda_{k_0}/(-m^2_{k_0})^{1/2}$
the dependence of the nucleation
rate on $k$ becomes more pronounced. We demonstrate this in the
second row of plots, for which
$\lambda_{k_0}=0.1\cdot k_0$,
$m^2_{k_0}=-0.013\cdot k_0^2$, $\gamma_{k_0}=-1.61 \cdot 10^{-3}~k_0^{3/2}$
and $k_f/k_0=0.0421$.
The dimensionless coupling is 
$\lambda_{k_0}/(-m^2_{k_0})^{1/2}$ $=0.88$ 
(instead of $0.48$ for the first row).
The first-order contribution 
$\ln ( A_k/k^4_f )$ in the expansion around the saddle point is
now comparable to the 
lowest order contribution $S_{k}$.
This 
indicates that higher orders in the expansion are important 
and the series stops converging. Therefore, there is a limit to
the validity of Langer's theory of homogeneous nucleation.

\section{Validity of the theory of homogeneous nucleation}
An intuitive derivation of the region of applicability of this theory
can be obtained through use of approximate analytical expressions for the 
pre-exponential factor. The dark triangles in the third plot of each 
row in fig.~\ref{fig1} display values predicted by the
approximate expression \cite{second,fifth}
\begin{equation} 
\ln \frac{A_k}{k^4}
\approx \frac{\pi k}{2} 
\left[
- \int_0^\infty \!\!\! r^3 \left[ 
U''_k\left( \phi_b(r) \right)
-U''_k\left( 0 \right)
\right] dr
\right]^{1/2}.
\label{eq:appr}
\end{equation}
Good agreement is observed with the numerical results denoted by 
stars, especially in the first case in which the expansion around the 
saddle point converges. 

Let us consider a potential given by eq.~(\ref{eq:two20}),
with $m^2_{k_0}>0$\footnote{
Models with $m^2_{k_0}>0$ and 
$m^2_{k_0}<0$ are related by a field shift \cite{second}. 
We have employed this shift in order to move the
metastable minimum to the origin in fig.~\ref{fig1}.}.
As the bubble-nucleation rate is largely independent of $k$ when
homogeneous nucleation theory is applicable, we can obtain an estimate 
of the rate by using $k=k_0$ and eq.~(\ref{eq:appr}).
Through appropriate rescalings of $r$ and $\phi$ we can define
a dimensionless potential 
$\tilde{U}(\phit)=\phit^2/2-\phit^3/3+h\,\phit^4/18$, with
$h=9\lx_{k_0} m_{k_0}^2/\gamma_{k_0}^2$. 
For $h \approx 1$ 
the two minima of the potential have approximately
equal depth, while for $h\to 0$ the
metastable minimum at the origin has negligible depth
compared to the stable one (spinodal point). 
One can estimate \cite{second}
\be
R=\frac{\ln \left(A_{k_0}/k_0^4 \right)}{S_{k_0}} \approx
T(h) \,\frac{\lx_{k_0}}{m_{k_0}}.
\label{fin} \ee
The function $T(h)$ diverges for $h\to 0$ \cite{second}, signalling the 
breakdown of homogeneous nucleation theory in the spinodal region.
Moreover, this breakdown may occur
even away from the spinodal region.
$T(h)$ is small for 
$h$ close to 1 \cite{second}, but a sufficiently large value of
$\lx_{k_0}/m_{k_0}$
can lead to $R \gta 1$.
This case corresponds to weak first-order phase transitions, as 
can be verified by observing that 
the saddle-point action,
the location of the true vacuum 
and the
difference in
free-energy density between the minima 
go to zero in the limit $m_{k_0}/\lx_{k_0}\to 0$ for fixed $h$. 

The above
conclusions are confirmed by the numerical computation of the nucleation rate.
A complete analysis of the region of validity of homogeneous nucleation
theory for potentials of the form of eq.~(\ref{eq:two20}) was presented
in ref.~\cite{second}. Also, two-scalar theories with potentials of various
forms were studied in ref.~\cite{third}. The conclusions can be summarized
as follows: Homogeneous nucleation theory relies on an expansion around the
saddle point that is assumed to dominate the tunnelling process (the critical
bubble). Therefore, it is applicable as long as this expansion converges.
Its breakdown is signalled by a pre-exponential factor comparable to the
leading exponential suppression and a strong dependence of the
predicted nucleation rate on the coarse-graining scale. 

\FIGURE{\epsfig{file=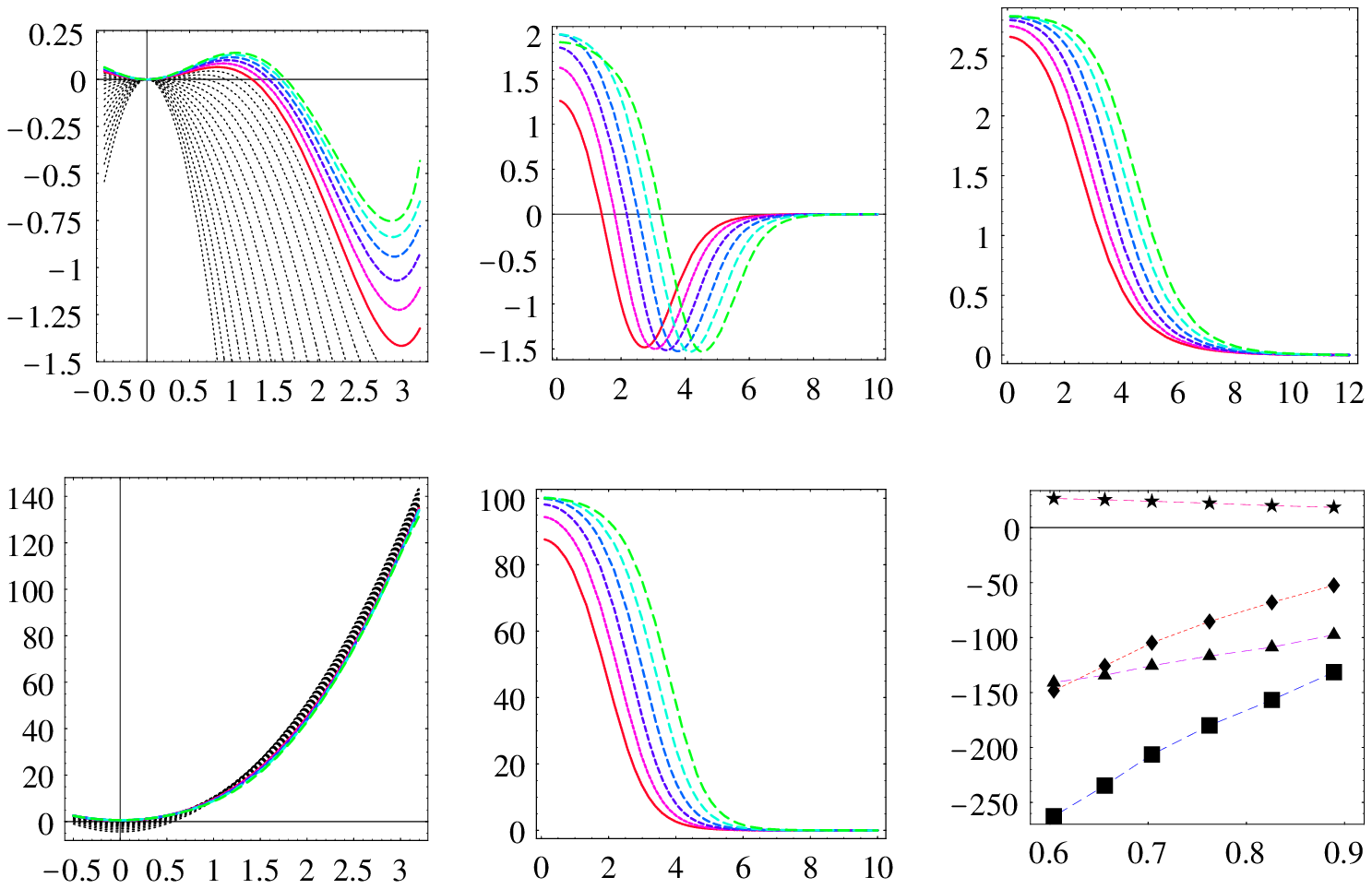,width=15.cm}%
        \caption{
A radiatively induced
first-order 
phase transition in a model with initial potential given by eq.~(\ref{pot3})
with $\phi^2_{0k_0}=1.712~k_0$, $\lx_{k_0}=0.01~k_0$ and
$g_{k_0}=0.2~k_0$.
The calculation of the bubble-nucleation rate is performed between the scales 
$k_i=e^{-4.7} k_0$ and $k_f=e^{-5.2} k_0$.
All dimensionful quantities are given in units of $k_f$.
}
	\label{fig2}}

\section{Radiatively induced first-order \\phase transitions}
These transitions 
\cite{colwein} form 
a class of special interest, as many cosmological phase transitions
belong to it. A typical example is presented in fig.~\ref{fig2} \cite{third}
for a two-scalar theory with an initial potential
\beq
U_{k_0}(\pha,\phb) =&& 
 \frac{\lx_{k_0}}{8}
\left[ ( \pha^2-\phi_{0k_0}^2 )^2 
+      ( \phb^2-\phi_{0k_0}^2 )^2 \right] 
\nonumber \\
&+& \frac{g_{k_0}}{4}  \pha^2 \phb^2,
\label{pot3} \eeq
with $\phi^2_{0k_0}=1.712~k_0$, 
$\lx_{k_0}=0.01~k_0$ and $g_{k_0}=0.2~k_0$.
In the first plot of the first row we depict a large part of the evolution of 
$U_k(\pha,\phb=0)$. 
The initial potential has only one minimum along the positive
$\pha$-axis and a maximum at the origin. 
In the sequence of potentials depicted by
dotted lines we observe the appearance of a new minimum
at the origin at some point in the evolution
(at $k_{CW}/k_0 \approx e^{-4.4}$). This minimum is generated by
the integration of fluctuations of the $\phb$ field, whose mass depends on
$\pha$ through the last term in eq.~(\ref{pot3})
(the Coleman-Weinberg mechanism \cite{colwein}). 
Immediately below we plot the mass term $\partial^2 U_k/\partial
\phb^2(\pha,\phb=0)$ of the 
$\phb$ field along the $\pha$-axis.
It turns positive at the origin at $k=k_{CW}$.
We calculate the nucleation rate using the potentials of the
last stages of the evolution. 
The solid lines correspond to $k_i/k_0=e^{-4.7}$, while the
line with longest dashes 
corresponds to $k_f/k_0=e^{-5.2}$. 

In the third plot of the first row we display the profile of the
saddle point $\phi_b(r)$ for various $k$ between $k_i$ and $k_f$.
In the middle row we plot the quantities
\be
W_{ik}(r)=\frac{\partial^2 U_k}{\partial \phi_i^2}\left(\pha=\phi_b(r)\right)-
\frac{\partial^2 U_k}{\partial \phi_i^2}\left(\pha=0\right)
\label{ww} \ee
for $\phb=0$ and $i=1,2$, i.e.
the difference of 
the mass terms of the two fields along the bubble profile and the false vacuum.
These quantities are expressed in units of $k_f$.
We observe that the mass of the $\phb$ fluctuations in
the interior of the critical bubble is much larger than the other
mass scales of the problem, which are comparable to $k_f$. 
This is a consequence of our choice of couplings $g_{k_0}/\lx_{k_0}=20$. 
Such a large ratio
is necessary for a strong first-order
phase transition to be radiatively induced. 

The results for the nucleation rate are summarized in the last plot
of the second row.
The horizontal axis corresponds to 
the ratio of the scale $k$
to  the mass of the $\pha$ field
at the absolute minimum. 
The dark diamonds give the negative of the action 
of the saddle point $-S_k$ at the scale $k$. 
The stars indicate the values of the prefactor  
$\ln ( A_{1k}/k^4_f )$ associated with 
$\pha$ fluctuations, while the triangles denote the prefactor
$\ln A_{2k}$ of the $\phb$ field.
A significant $k$ dependence of all these quantities is observed. 
The dark squares give our results for 
$\ln(I/k^4_f ) 
= -S_k+\ln [ A_{1k} A_{2k}/k^4_f ]$. 
This quantity has a strong 
$k$ dependence, which signals the breakdown of the 
expansion around the saddle point. 
This is expected, as
$|\ln A_{2k}|$ is comparable to the
saddle-point action $S_k$, even though $\ln ( A_{1k}/k^4_f )$
remains small. 

The breakdown of the expansion is a generic problem of all the 
radiatively induced first-order phase transitions we have studied
\cite{third}. It can be understood in physical terms by considering
the quantity $W_{2k}(r)$ defined in eq.~(\ref{ww}) and depicted in the
second plot of the second row in fig.~\ref{fig2}. Due to the large values
of the ratio $g_{k_0}/\lx_{k_0}=20$, the mass of the $\phb$ field is
much larger than $k_f$, the typical scale of the tunnelling
process. This indicates that fluctuations in the $\phb$ direction
cost excessive amounts of free energy and are suppressed. As these
fluctuations are inherent to the system, the total nucleation rate
should be suppressed as well. This explains the very negative
values of $\ln A_{2k}$. 
This behaviour is not surprising. The radiative corrections to the 
potential and the pre-exponential factor have a very similar form 
of fluctuation determinants. When the radiative corrections are large enough
to modify the initial potential and generate a new minimum, the 
pre-exponential factor should be expected to be important also. 

We point out that the presence of a region of negative values for the
function $W_{1k}(r)$, depicted in the 
second plot of the first row in fig.~\ref{fig2}, results in 
the existence of a class of $\pha$ fluctuations with small typical 
free energy. As a result, deformations of the critical bubble in
the $\pha$ direction are favourable and their effect tends to enhance
the total nucleation rate. This explains the positive
values of $\ln ( A_{1k}/k^4_f )$.

Finally, we mention that the prefactor $A_{2k}$ can be estimated using
approximate expressions analogous to eq.~(\ref{eq:appr}) and appropriate
rescalings. One finds \cite{fifth}
\be
\frac{\ln  A_{2k_f}}{S_{k_f}} \approx
- R(h) \,\frac{\lx_{k_0}}{g_{k_0}},
\label{finn} \ee
with $R(h) \gta 24$ for any $h$. 
This expression indicates that only very small values of 
$\lx_{k_0}/g_{k_0}$ may lead to a convergent expansion 
around the saddle point. However, eq.~(\ref{finn})
is not valid for $\lx_{k_0}$ smaller than $g_{k_0}$ by
more than two orders of magnitude. We have not found any 
model for which the numerical study of a radiatively induced
first-order phase transition indicates a convergent expansion.

\section{Two dimensions}

\FIGURE{\epsfig{file=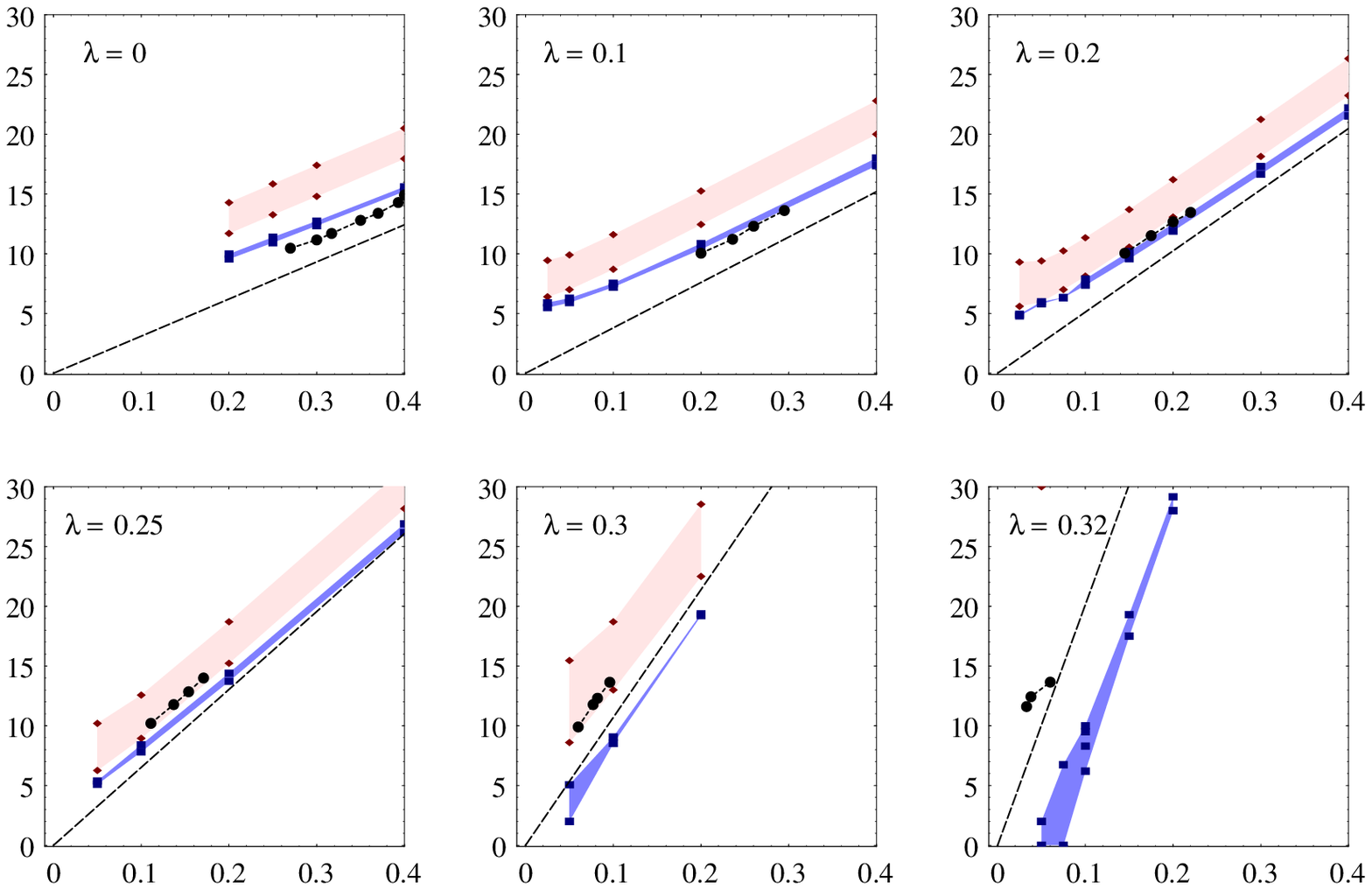,width=15.cm}%
        \caption{
Comparison of our method with lattice studies:
Diamonds denote the saddle-point action $S_k$ 
and squares the bubble-nucleation rate 
$-\ln\left(I/m^3\right)$ 
for $k=1.2\,k_f$ and $2\,k_f$. 
Dark circles denote the results for the nucleation rate from
the lattice study of ref.~\cite{lattice}. 
}
	\label{fig3}}

A crucial confirmation of the reliability of our approach was obtained
in ref.~\cite{fourth},
where bubble nucleation was studied for (2+1)-dimensional theories. 
Our results were compared with those of the lattice simulations 
of ref.~\cite{lattice}. We considered potentials similar to the
one of eq.~(\ref{eq:two20}), with small modifications in order to
match the continuum theory of the lattice simulations. 
The technical aspects of the application of our formalism to two
dimensions can be found in ref.~\cite{fourth}. 

In fig.~\ref{fig3} 
we present a comparison of results obtained through our method
with lattice results.
The dimensionless parameter $\lx$ is approximately equal to 
$3 \lx_{k_0}/(\theta\, m^2_{k_0})$ in terms of the parameters of the
potential of eq.~(\ref{eq:two20}). The dimensionless
variable $\theta$ (that plays the 
role of a rescaled temperature)
is approximately equal to $\gamma^2_{k_0}/m^4_{k_0}$. The values
of $1/\theta$
are given on the horizontal axis of the plots in fig.~\ref{fig3}. 
(For the details, see ref.~\cite{fourth}.)
The diamonds denote the saddle-point action $S_k$. 
For every choice of $\lx$, $\theta$ we determine $S_k$
at two scales: $1.2\,k_f$ and $2\,k_f$. The light-grey region between
the corresponding points gives an indication of the
$k$ dependence $S_k$.
The negative of the 
bubble-nucleation rate $-\ln\left(I/m^3\right)$ is denoted 
by dark squares. 
The dark-grey region between the values obtained at 
$1.2\,k_f$ and $2\,k_f$ gives a good check of the convergence of the
expansion around the saddle-point. If this region is thin, the
prefactor is in general small and
cancels the $k$ dependence of the action.
The dark circles denote the results for the nucleation rate from
the lattice study of ref.~\cite{lattice}. 

For $\lx=0$, $0.1$, $0.2$
the values of $-\ln(I/m^3)$ computed at $1.2\,k_f$ and $2\,k_f$
are equal to a very good approximation. This
confirms the convergence of the expansion around the saddle-point and 
the reliability of the calculation. The $k$ dependence of
the saddle-point action is cancelled by the prefactor, so that the 
total nucleation rate is $k$ independent. Moreover, the
prefactor is always significantly smaller than the
saddle-point action.  
The agreement with the lattice predictions is 
good. It is clear that the contribution of
the prefactor is crucial for the correct determination of the
total bubble-nucleation rate. 

For larger values of $\lx$ 
the matching between the lattice and the renormalized actions
becomes imprecise. As a result, we cannot make sure that we are studying the
same theory as the one simulated on the lattice. 
This generates deviations of our results from the lattice ones, which
start becoming apparent for $\lx=0.25$.
However,
the internal consistency criteria of our method
provide a test of the convergence of the expansion
around the saddle point in all cases.
The consistency of our calculation is achieved  
for $1/\theta \gta 0.12$ even for $\lx=0.32$.
However, 
the breakdown of the expansion is apparent
for $1/\theta \lta 0.12$.
The prefactor becomes comparable to the saddle-point
action and the
$k$ dependence of the predicted nucleation rate is strong.

\section{Applications}

As a first example we consider the
bound on the Higgs-boson mass from the stability of our vacuum.
The top-quark radiative corrections to the zero-temperature effective
potential of the Higgs field may result in the appearance of a new
minimum, deeper than the one located at 247 GeV. In order for this not to
happen, a lower bound on the Higgs-boson mass must be imposed.
This bound can be relaxed if one allows for the presence of a new vacuum,
but demands that the time needed for a transition to
it is larger than the age of our universe. The largest 
rates are associated with thermal transitions at high temperatures, from
a metastable minimum in the symmetric phase of the Standard Model directly to 
the absolute minimum at very large Higgs-field expectation values 
\cite{anderson}. 

We consider 
the effective three-dimensional description of
the top-Higgs system at a scale $k_0\approx T$. 
The top-quark 
fluctuations are almost decoupled because of the absence of a
zero Matsubara frequency for fermions.
This permits a very simple determination of the potential $U_{k_0}(\phi)$, 
which is 
equal to the sum of the zero-temperature potential and
the contribution from the top-quark thermal fluctuations \cite{fifth}.
It can be written as 
\be
U_{k_0}(\phi) \approx \frac{m^2}{2} \phi^2 - \frac{\kx}{4}  \phi^4,
\label{toppot} \ee
where
$m^2 = g^2_{t4}\, T^2/4$, 
$\kx = |\lx_4| \, T$, $g_{t4}$ is the top-quark Yukawa coupling and
$\lx_4$ the average (negative) value of the quartic coupling 
over the region of Higgs-field expectation values
relevant for the critical-bubble profile.
The term $\sim \phi^2$ is
the leading temperature
correction in the expansion of the one-loop 
contribution from the top quark. 

The action of the 
saddle point is \cite{fifth} 
\be
S_{k_0} \approx 18.9 \frac{m}{\kx} = 9.45 \, 
\frac{g_{t4}}{|\lx_4|}.
\label{action1} \ee
The prefactor can be estimated through eq.~(\ref{fin}) 
\be
\ln  \left( A_{k_0}/k^4_0 \right)  
\approx 2.00 \, \frac{k_0 \, \pi}{2 \, m}=
\frac{6.28}{g_{t4}}.
 \label{prefactor1} \ee
For the experimental value of the top-quark mass ($g_{t4} \approx 1$) we find
$|\lx_4|\lta 0.05$ for all Higgs-field expectation values, 
in agreement with ref.~\cite{anderson}. 
As a result, the prefactor is expected to give only a small correction to the 
bubble-nucleation rate. 

The electroweak phase transition is 
radiatively induced by gauge field fluctuations.
Even though the generalization of our formalism to gauged systems
requires further work, an estimate of the validity of the expansion
around the saddle point can be obtained by exploiting the resemblance
between the gauge-field fluctuation determinants and the ones in two-scalar
models. 
The estimate for the prefactor $|{\ln  A_{2k_f} }|$, associated
with the gauge fields,
is given by eq.~(\ref{finn}), with
$\lx/g$ taking the ``effective'' value \cite{fifth}
\be
\left(\frac{\lx}{g}\right)_{\!\!\rm SM} \approx 
\frac{(4 m_W+2 m_Z)\,m^2_H}{
4( 4 m^3_W+2 m^3_Z)}=0.35 \, \left( \frac{m_H}{{\rm 100 \,GeV}}\right)^2.
\label{b1} \ee
For the prefactor to be smaller than
1/2 of the action, 
$m_H \lta 25$ GeV is required. 

Higgs-boson masses below the experimental lower limit of 90 GeV 
are of academic interest. Moreover, 
for $m_H \gta 80$ GeV, there is
no phase transition in the high-temperature Standard Mo\-del.
There is the possibility, however, for a sufficiently strong 
first-order phase transition for ba\-ryogenesis within the Minimal
Supersymmetric Standard Model with a very 
light stop~\cite{mssm}. 
In this model, the ``effective'' value of $g/\lx$ obeys \cite{fifth}
\begin{eqnarray}
\left(\frac{\lx}{g}\right)_{\!\!\rm MSSM} &\gta &
\frac{(4 m_W+2 m_Z+6 m_t)\,m^2_H}{
4( 4 m^3_W+2 m^3_Z+6 m^3_t)} \nonumber\\
&=&0.11 \, \left( \frac{m_H}{{\rm 100 \,GeV}}\right)^2.
\label{b2} 
\end{eqnarray}
For $m_t= 175$ GeV,
in order to have a prefactor 
smaller than
1/2 of the action, $m_H\lta 45$ GeV is required.


\end{document}